\documentstyle[epsfig]{aipproc}
\begin{document}
\title{A Strong Electroweak Sector \\ at Future $\mu^+\mu^-$ Colliders
\thanks{
Talk presented by D. Dominici. The research of RC, SDC, DD, AD and RG has 
been carried out within the Program
Human Capital and Mobility: ``Tests of electroweak symmetry
breaking and future European colliders'', CHRXCT94/0579.
JFG is supported by the U.S. Department of Energy
under grant No. DE-FG03-91ER40674
and by the U.C. Davis Institute for High Energy Physics. JFG would
like to thank V. Barger, M. Berger and T. Han for collaboration on
Higgs discovery at a muon collider, as part of which project many
of the techniques employed in the $P^0$ discussion were developed.}}
\author{R. Casalbuoni$^{a,b,d}$, S. De Curtis$^b$, D. Dominici$^{a,b}$}
\author{A. Deandrea$^c$, R. Gatto$^d$ and J. F. Gunion$^e$}
\address{$^a$Dipartimento di Fisica Universit\`a di Firenze,
I-50125 Firenze, Italia\\ $^b$ I.N.F.N., Sezione di Firenze,
I-50125 Firenze, Italia\\ $^c$Centre de Physique Th\'eorique,
CNRS, Luminy F-13288 Marseille, France\\ $^d$D\'epartement de
Physique Th\'eorique, Universit\'e de Gen\`eve, CH-1211 Gen\`eve
4, Suisse\\ $^e$ Department of Physics, University of California,
Davis, CA 95616, USA}

\maketitle

\def\MPL #1 #2 #3 {{\sl Mod.~Phys.~Lett.}~{\bf#1} (#3) #2}
\def\NPB #1 #2 #3 {{\sl Nucl.~Phys.}~{\bf B#1} (#3) #2}
\def\PLB #1 #2 #3 {{\sl Phys.~Lett.}~{\bf B#1} (#3) #2}
\def\PR #1 #2 #3 {{\sl Phys.~Rep.}~{\bf#1} (#3) #2}
\def\PRD #1 #2 #3 {{\sl Phys.~Rev.}~{\bf D#1} (#3) #2}
\def\PRL #1 #2 #3 {{\sl Phys.~Rev.~Lett.}~{\bf#1} (#3) #2}
\def\RMP #1 #2 #3 {{\sl Rev.~Mod.~Phys.}~{\bf#1} (#3) #2}
\def\ZPC #1 #2 #3 {{\sl Z.~Phys.}~{\bf C#1} (#3) #2}
\def\IJMP #1 #2 #3 {{\sl Int.~J.~Mod.~Phys.}~{\bf#1} (#3) #2}

\def\lsim{\mathrel{\raise.3ex\hbox{$<$\kern-.75em\lower1ex\hbox{$\sim$}}}}
\def\gsim{\mathrel{\raise.3ex\hbox{$>$\kern-.75em\lower1ex\hbox{$\sim$}}}}
\def\sigrts{\sigma_{\tiny\rts}^{}}
\def\sigrtssq{\sigma_{\tiny\rts}^2}
\def\sigrtsprime{\sigma_{E}}
\def\nsigrts{n_{\sigrts}}
\def\mupmum{\mu^+\mu^-}
\def\lplm{\ell^+\ell^-}
\def\drts{\Delta\sqrt s}
\def\rts{\sqrt s}
\def\ie{{\it i.e.}}
\def\eg{{\it e.g.}}
\def\eps{\epsilon}
\def\anti{\overline}
\def\wp{W^+}
\def\wm{W^-}
\def\mw{m_W}
\def\mz{m_Z}
\def\fbi{~{\rm fb}^{-1}}
\def\fb{~{\rm fb}}
\def\pbi{~{\rm pb}^{-1}}
\def\pb{~{\rm pb}}
\def\mev{~{\rm MeV}}
\def\gev{~{\rm GeV}}
\def\tev{~{\rm TeV}}

\def\pzero{P^0}
\def\mpzero{m_{\pzero}}
\def\pzerop{P^{0\,\prime}}
\def\gs{g^{\prime\prime}}
\newcommand{\be}{\begin{equation}}
\newcommand{\ee}{\end{equation}}
\newcommand{\bea}{\begin{eqnarray}}
\newcommand{\eea}{\end{eqnarray}}
\newcommand{\nn}{\nonumber}
\begin{abstract}
We discuss the prospects for detecting at a muon collider
the massive new vector resonances $V$
and light pseudo-Nambu-Goldstone bosons $P$ of a typical strongly interacting
electroweak sector (as represented by the BESS model). Expected sensitivities
to $V$'s at a high energy collider are evaluated and the excellent
prospects for discovering $P$'s via scanning at a low energy collider
are delineated.
\end{abstract}

\section*{Introduction}

In this contribution we consider  some aspects of
strong electroweak symmetry breaking at future $\mu^+\mu^-$
colliders. We will concentrate on the possibility
of detecting new vector resonances and pseudo-Nambu-Goldstone
bosons (PNGB) originating from the strong interaction responsible
for electroweak symmetry breaking.
 The importance of
technivector and technipion physics at muon
colliders  was discussed during the strong dynamics
working group meetings \cite{elw}.
This study will be here
performed within the framework of the BESS model \cite{bess} and
of its generalizations \cite{bessu8}. We recall  the main
features of this model. The BESS model is an effective lagrangian
parameterization of the symmetry breaking mechanism, based on a
symmetry $G=SU(2)_L\otimes SU(2)_R$ broken down to $SU(2)_{L+R}$.
New vector particles are introduced as gauge bosons associated with
a hidden $H'=SU(2)_V$. The symmetry group of the theory becomes
$G'=G\otimes H'$. It breaks down spontaneously to $H_D=SU(2)$,
which is the diagonal subgroup of $G'$. This gives rise to six
Goldstone bosons. Three are absorbed by the new vector particles
while the other three give mass to the standard model (SM) gauge
bosons, after the gauging of the subgroup $SU(2)_L\otimes
U(1)_Y\subset G$. The parameters of the BESS model are  the mass
of these new bosons $M_V$, their self coupling $\gs$,  and a
third parameter $b$ whose strength characterizes the direct
couplings of the new vectors $V$ to the fermions. However, due to
the mixing of the $V$ bosons with $W$ and $Z$, the new particles
are coupled to the fermions even when $b=0$. The parameter $\gs$
is expected to be large due to the fact that these new gauge
bosons are thought of as bound states from a strongly interacting
electroweak sector. By taking the formal $b\rightarrow 0$ and
$\gs\rightarrow\infty$ limits, the new bosons decouple and the SM
is recovered. By considering only the limit
$M_V\rightarrow\infty$ they do not decouple.

The extension of the BESS model we will consider here is obtained
by enlarging the original chiral symmetry $SU(2)_L\otimes
SU(2)_R$ to the larger group $SU(8)_L\otimes SU(8)_R$
\cite{bessu8}. The main new feature of this extension is the presence
of 60 PNGB's. Their masses come from the breaking
of the chiral group provided by the $SU(3)\otimes SU(2)_L\otimes
U(1)_Y$ gauge interactions and by Yukawa couplings \cite{mass}.
We emphasize that most non-minimal models of a strongly interacting
electroweak sector will contain PNGB's, although
the number and their exact properties are model dependent.

\section*{Bounds on the parameter space for the new vector bosons}

\begin{figure} 
\vspace{10pt}
\centerline{\epsfig{file=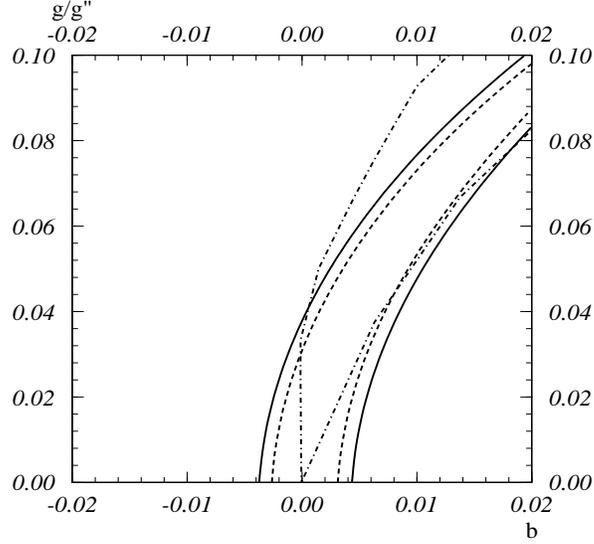,height=3.0in,width=3.5in}}
\vspace{10pt}
\caption{BESS model 90\% C.L. contours in the
 plane $(b,g/\gs)$ for $M_V=1\tev$.
The continuous and the dashed lines correspond to the bound from
the differential $W_{L,T}W_{L,T}$ cross sections at a $500\gev$
lepton collider for $B=0.1$ and $B=0.2$ respectively. The
dot-dashed line corresponds to the bound from the total cross
section $pp\to W^\pm,V^\pm\to W^\pm Z$ at LHC. The allowed
regions are  the internal ones.}
\label{fig1}
\end{figure}

Bounds on the parameter space of the  BESS model from existing
data have already been studied (see for instance \cite{desy}).
Future lepton colliders can improve these limits by testing
virtual effects of the new vector particles, especially in the
annihilation channels $l^+l^-\to W^+W^-$. In fact, the most
relevant observable is the differential cross section $ d\sigma (
l^+l^-\to W^+_{L,T}W^-_{L,T})/d\cos\theta$, where $\theta$ is the
overall center of mass scattering angle  and the decays of the $W^+$ and
$W^-$ are used to reconstruct the final $W$ polarizations. The
analysis is performed by taking 19 bins in the angular region
restricted by $\vert \cos\theta\vert\leq 0.95$. Since the new
vectors strongly couple to the longitudinal $W$, the most relevant
process is $l^+l^-\to W^+_LW^-_L$. We have studied the channel
with one $W$ decaying leptonically and the other one
hadronically. In Fig. \ref{fig1} we present the
 90\% C.L. contours in the  plane $(b,g/\gs)$ for $M_V=1\tev$.
The continuous and the dashed lines correspond to the bound from
the combined differential $W_{L,T}W_{L,T}$ cross sections at a
$500\gev$ lepton collider with $ 20~fb^{-1}$ of integrated
luminosity, assuming effective branching ratios $B=0.1$ and
$B=0.2$, respectively. The first case can correspond to an
$e^+e^-$ machine where the loss of luminosity from beamstrahlung
is taken into account. The second one to a muon collider or to an
electron collider with a low beamstrahlung loss.
 The regions within which one can exclude or detect
the $V$ at the 90\% CL are  the external ones.

Statistical errors are taken into account and we have assumed a
systematic error of $1.5\%$.  The dot-dashed line corresponds to the
bound from the total cross section $pp\to W^\pm,V^\pm\to W^\pm
Z\to\mu\nu\mu^+\mu^-$ at LHC  if no deviation is observed with
respect to the SM within the experimental errors. Statistical 
errors and a systematic error of $5\%$ have been included. In conclusion,
for models of new strong interacting vector resonances
the measurement of $d\sigma( l^+l^-\to
W^+_{L,T}W^-_{L,T})/d\cos\theta $ gives rather strong bounds, provided
one is able to reconstruct the final $W$ polarizations.
These bounds become very stringent for increasing energy of the
collider \cite{desy}.

Direct production  of the
new vector bosons can also be considered. A muon collider of
$3-4\tev$ will enable us to completely explore the
strongly interacting electroweak option \cite{gunion4T}.

Since one  must rely on an effective non-renormalizable
description, one has to take into account the partial wave
unitarity limits from $WW$ scattering. In fact when the mass of
the new vectors is in the range $2-4\tev$, these bounds come out
to be quite restrictive. If we denote  by $A(s,t,u)$ the
amplitude for the scattering $W^+W^-\to ZZ$, one gets
\be
A(s,t,u)= \left(1-\frac 3 4\alpha\right)\frac s{v^2}+\frac \alpha
4 \frac{M_V^2}{v^2}\left(\frac {u-s} {t-M_V^2+iM_V\Gamma_V}+
\frac{t-s}{u-M_V^2+iM_V\Gamma_V}\right)
\ee
where $\alpha=192 \pi v^2\Gamma_V/M_V^3$ and  $v=246\gev$.

Projecting the components with definite isospin into the lower
partial waves and requiring that the bounds $a_{IJ}\le 1$ are
valid up to energies $\Lambda$, such that $\Lambda/M_V\le 1.5$,
we get the limitations  in the plane $(M_V,\Gamma_V)$ given in
Fig. 2. The intersection of the three allowed regions gives a
general upper bound on the mass $M_V\sim 3\tev$.
The previous considerations can be applied also to the technirho
case, which is obtained by taking $\alpha=2$. In this model the
unitarity bound turns out to be $M_{\rho_T}\le 2$ TeV.
In Fig. 3 we translate these
limits into restrictions on the parameters of the BESS model, using the
relation $\Gamma_V=M_V^5/(48\pi v^4 {\gs}^2)$.

We conclude that the unitarity bounds imply that one or more of the
heavy vector resonances should be discovered at the LHC, NLC,
a $\rts\sim 500\gev$ muon collider, or, for certain, 
at a $3-4\tev$ muon collider, unless $\gs$ is very large
and $b$ is very small so that they are largely decoupled.

\begin{figure} 
\vspace{10pt}
\centerline{\epsfig{file=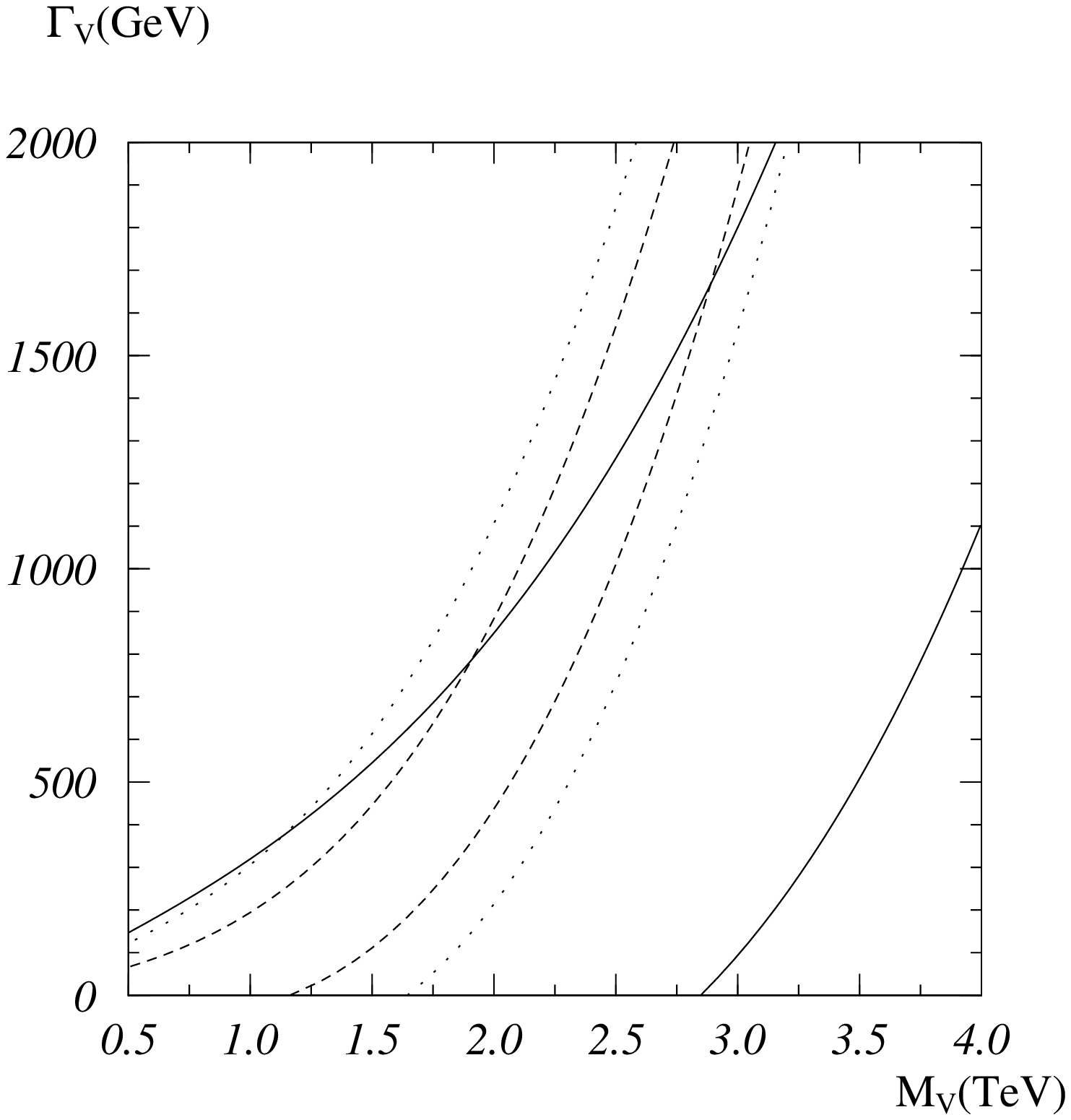,height=3.0in,width=3.5in}}
\vspace{10pt}
\caption{Unitarity bounds in the plane $(M_V, \Gamma_V)$ with
$\Lambda/M_V=1.5$. The dashed line corresponds to the bound from
the partial wave $a_{00}$, the dotted one to $a_{20}$, and the
continuous one to $a_{11}$. The allowed regions are the internal
ones.}
\label{fig2}
\end{figure}

\begin{figure} 
\vspace{10pt}
\centerline{\epsfig{file=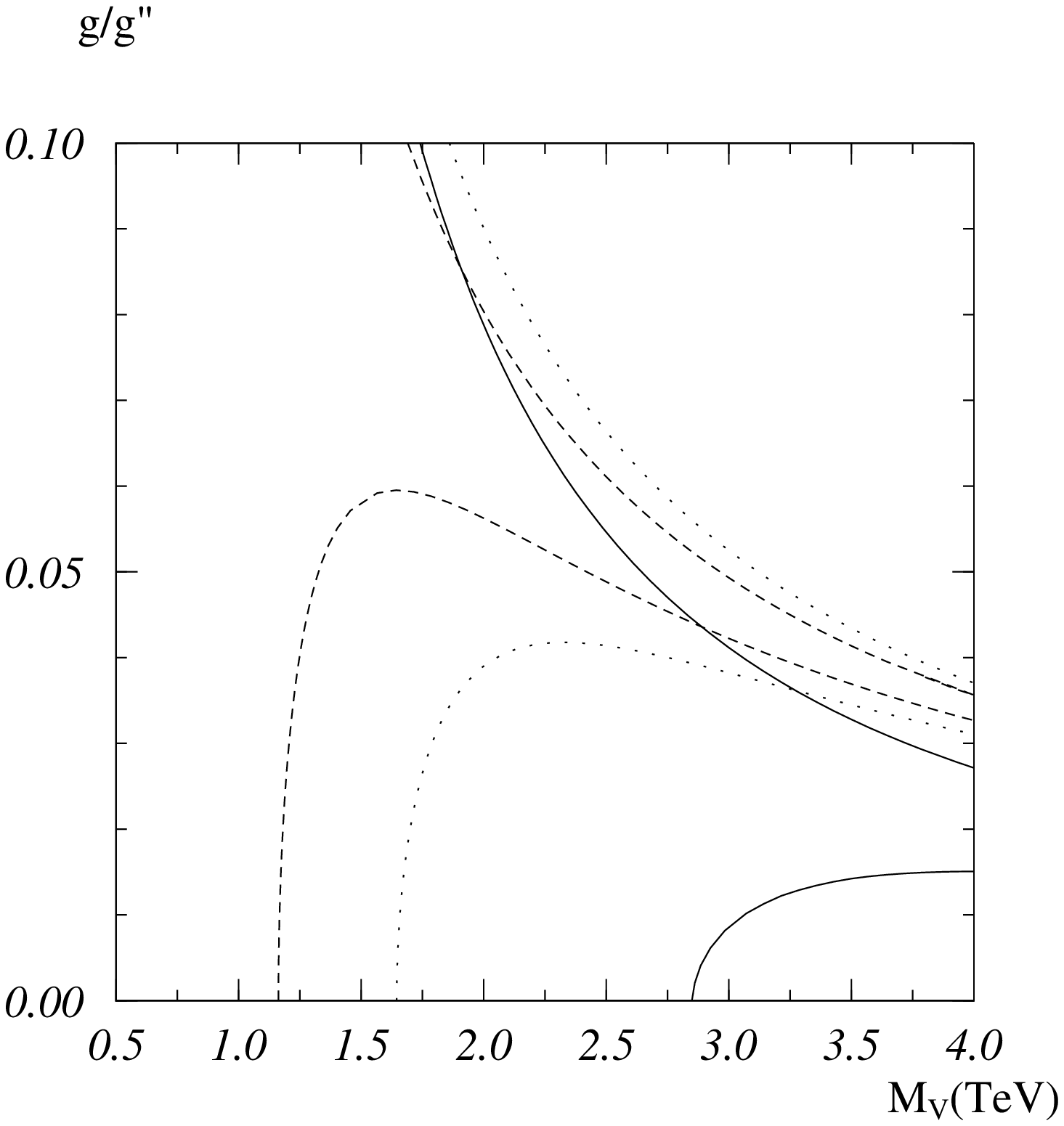,height=3.0in,width=3.5in}}
\vspace{10pt}
\caption{Unitarity bounds in the plane $(M_V, g/\gs)$ with
$\Lambda/M_V=1.5$. The dashed line corresponds to the bound from
the partial wave $a_{00}$, the dotted one to $a_{20}$, and the
continuous one to $a_{11}$. The allowed regions are the internal
ones.}
\label{fig3}
\end{figure}

\section*{PNGB production at a muon collider in the
extended BESS model}

In this section, we consider $s$-channel production of the lightest
neutral PNGB $\pzero$ at a future $\mu^+\mu^-$ collider.
  Although
we shall employ the specific $\pzero$ properties as predicted
by the extended BESS model with $SU(8)\otimes SU(8)$
symmetry \cite{bessu8}, many of our results apply in general fashion
to other models of a strongly interacting electroweak sector.
An example of large cross section for the production of  
an isoscalar and
an isovector technipion of $110~GeV$
 was shown by Bhat \cite{pushpa}.

In the extended BESS model, the PNGB mass 
derives   both from gauge contributions and 
from the effective low-energy Yukawa interactions 
between the PNGB's and ordinary fermions \cite{mass}. 
The lightest neutral PNGB's are the following combinations of the isosinglet
and isotriplet components:
$\pzero={(\tilde\pi_3-\pi_D)}/{\sqrt{2}}$, 
$\pzerop={(\tilde\pi_3+\pi_D)}/{\sqrt{2}}$.
The  $\pzero$ boson couples to the $T_3=-1/2$ component of the fermion
doublet while  $\pzerop$ couples to the $T_3=1/2$ component.
It is the $\pzero$ upon which we focus. 
The expressions for the $\pzero$ and $\pzerop$ masses are
\cite{mass}
\be
m^2_{\pzero}=\frac{2\Lambda^2}{\pi^2v^2} m_b^2,~~~~
{m^2_{\pzerop}}=\frac{2\Lambda^2}{\pi^2v^2} m_t^2
\ee
where $\Lambda$ is an $UV$ cut-off, situated in the $TeV$ region.
The first result above can be written as $\mpzero\sim 8\gev \times\Lambda({\rm
TeV})$.
Thus, not only does the $\pzero$ have the
$\mu^+\mu^-$ coupling needed for $s$-channel production at a muon
collider, but also, in this model, $\mpzero$ should be relatively
small, $\lsim 80\gev$ for $\Lambda\lsim 10\tev$ (as expected in the
present model).

\begin{figure}[b] 
\vspace{10pt}
\centerline{\epsfig{file=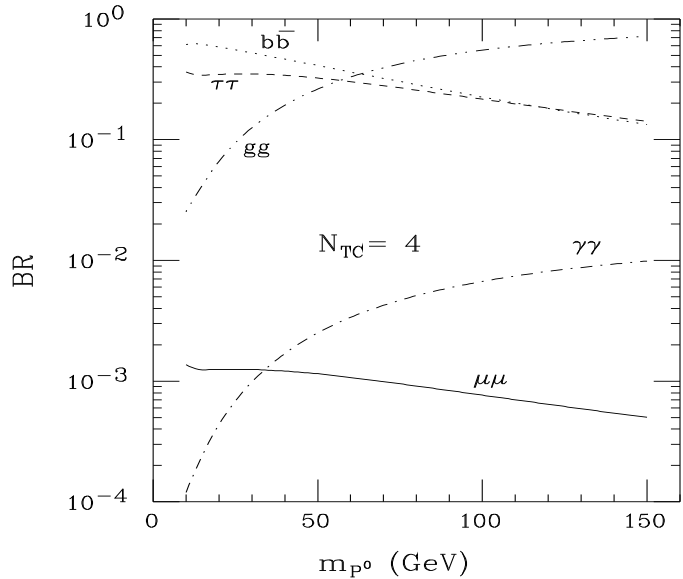,height=3.0in,width=3.5in}}
\vspace{10pt}
\caption{Branching ratios for $\pzero$ decay into $\mu^+\mu^-$,
$\tau^+\tau^-$, $b\bar b$, $\gamma\gamma$,
and $gg$.}
\label{figbrs}
\end{figure}

The $\pzero$ Yukawa couplings to fermions are \cite{mass}
\be
{\cal L}_Y= -i \lambda_b \bar b\gamma_5 b \pzero
-i \lambda_\tau \bar \tau\gamma_5 \tau \pzero
-i\lambda_\mu\bar\mu\gamma_5\mu \pzero
\ee
with
\be
\lambda_b=\sqrt{\frac 2 3 }\frac {m_b} v,~~~
\lambda_\tau=-\sqrt{6}\frac {m_\tau} v,~~~
\lambda_\mu=-\sqrt{6} \frac {m_\mu} v
\label{pcoups}
\ee
For the $\pzero$, the $\gamma\gamma$ and
gluon-gluon channels are also important; the corresponding 
couplings are generated by the ABJ anomaly.
Clearly these couplings are model dependent and, as an example, we
borrow them from technicolor theories \cite{anom}.  We
list here all the partial widths relevant for our analysis:
\bea
\Gamma(\pzero\to \bar f f) &=& C \frac {m_{\pzero}}{8\pi} \lambda_f^2 (1-\frac
{4 m_f^2}{m^2_{\pzero}})^{1/2}\nn\\
\Gamma (\pzero\to gg)&=& \frac {\alpha_s^2}{48\pi^3 v^2} N^2_{TC}m^3_{\pzero}\nn\\
\Gamma (\pzero\to\gamma\gamma)&=& \frac {2\alpha^2}{27\pi^3 v^2} N^2_{TC}m^3_{\pzero}
\eea
where $C=1(3)$ for leptons (down-type quarks) and $N_{TC}$ is the number of
technicolors. The corresponding branching ratios are shown in
Fig. \ref{figbrs}. 

There are presently no
definitive limits on the mass of the $\pzero$.  Potentially useful
production modes arise through its ABJ anomaly coupling to pairs of
electroweak gauge bosons \cite{randa,chivu}. At LEP the dominant
production mode is $Z\to\gamma \pzero$. The limit
of \cite{chivu}, obtained by requiring a $Z\to\gamma \varphi$
decay width of $2 ~10^{-6}\gev$ in order to make the $\varphi$ visible
in a sample of $10^7$ $Z$ bosons, can be rescaled to the case of the $\pzero$.
We find that for $N_{TC}\le 9$ there is no limit on
$m_{\pzero}$, while, for instance, for $N_{TC}=10$ $m_{\pzero}\ge 12\gev$
is required.

Future limits from the Tevatron and LHC have also been considered
\cite{chivu}. In the single production mode, the best hope of
finding the $\pzero$ at hadron colliders is via the anomalous decay
$\pzero\to\gamma\gamma$. The signal in this channel is similar
to that of a  Standard Model Higgs boson of the same mass, 
given the comparable branching ratio illustrated in Fig.~\ref{figbrs}.
However, for the range of $\pzero$ masses we
are considering the signal will be hard to see since the
$\gamma\gamma$ continuum background is very large at low mass.
Another possibility would be to produce pairs of PNGB's, as for
instance, in the resonant production $pp\to V^\pm\to P^\pm \pzero+X$
\cite{PGB}, where $V$ is the vector resonance discussed in the
Introduction. However, the discovery of the PNGB's via $\bar t b
\bar b b$ or $\bar t b gg$ decays, needs a careful evaluation of
backgrounds in the LHC environment. One could also consider the
process $pp\to gg\to \pzero\pzero$, mediated by the anomalous $gg\pzero$ vertex,
which could be detected by looking for equal mass pairs. Again,
backgrounds will be large. Thus, as far as we know, 
reliable bounds will not be obtained at hadron colliders.

\begin{figure}[h] %
\vspace{10pt}
\centerline{\epsfig{file=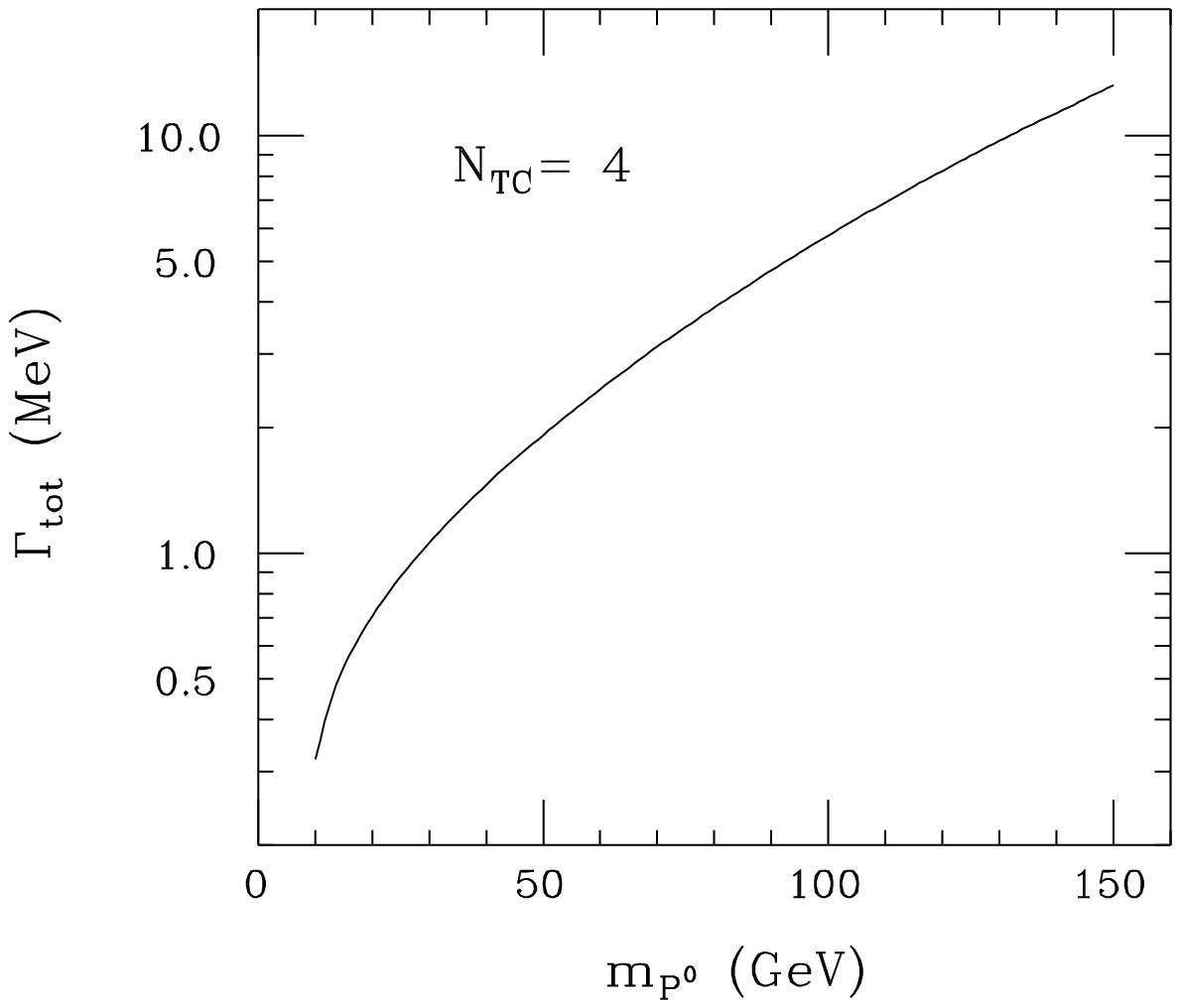,height=3.0in,width=3.5in}}
\vspace{10pt}
\caption{$\Gamma_{\rm tot}$ for the $\pzero$ as a function of $\mpzero$.}
\label{figgamtot}
\end{figure}

Thus, it is clearly of great importance to find a means for 
discovering or eliminating a $\pzero$ with mass
between, roughly, $10\gev$  and $100\gev$.
For much of this mass range a muon collider would be the ideal probe.
First, we note that the $\pzero$ has a sizeable $\mupmum$ coupling
[eq.~(\ref{pcoups})]. Second,
the muon collider is unique in its ability to achieve 
the very narrow Gaussian spread, $\sigrts$, in $\rts$
necessary to achieve a large $\pzero$ cross section given
the very narrow width of the $\pzero$ (as plotted in
Fig.~\ref{figgamtot}). One can achieve $R=0.003\%$ beam energy resolution
with reasonable luminosity at the muon collider, leading to
\begin{equation}
\sigrts\sim 1\mev\left({R\over 0.003\%}\right)\left({\rts\over
50\gev}\right)\,;
\label{sigrtsdef}
\end{equation}
in addition, 
the beam energy can be very precisely tuned ($\Delta E_{\rm beam}\sim
10^{-5}E_{\rm beam}$ is `easy'; $10^{-6}$ is achievable) as crucial
for scanning for a very narrow resonance.

\begin{figure}[b] 
\vspace{10pt}
\centerline{\epsfig{file=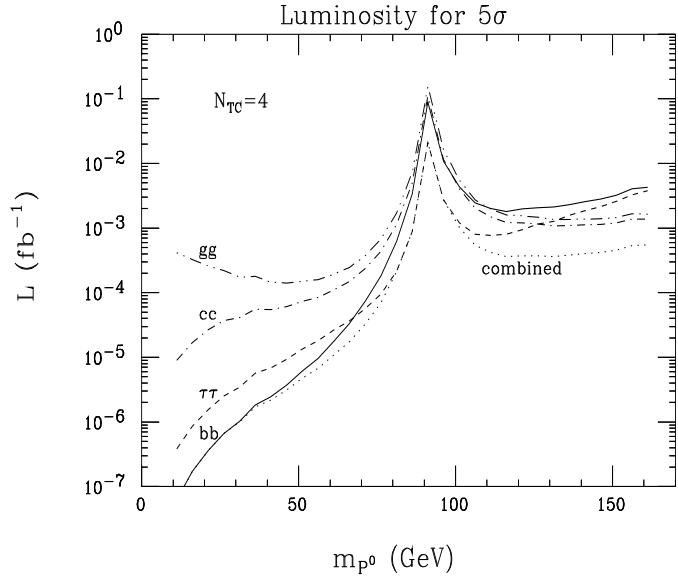,height=3.0in,width=3.5in}}
\vspace{10pt}
\caption{Luminosity for a 5$\sigma$ signal of $\pzero$ for the channels:
 $b\bar b$, $\tau^+\tau^-$, $c\bar c$, $gg$, and for the optimal combination
of these four channels.}
\label{figlum}
\end{figure}

To quantitatively assess the ability of the muon collider to discover
the $\pzero$ we have proceeded as follows. We compute the $\pzero$
cross section by integrating over the resonance using a $\rts$
distribution given by a Gaussian of width $\sigrts$ (using $R=0.003\%$)
modified by bremsstrahlung photon emission. 
(Beamstrahlung is negligible at a muon collider.) See Ref.~\cite{bbgh}
for more details. We separate $\tau^+\tau^-$, 
$b\anti b$, $c\anti c$ and $q\anti q$, $ gg$
final states by using topological and $\tau$ tagging with efficiencies 
and mistagging probabilities as estimated by B. King \cite{bking}:
$\eps_{bb}=0.55$, $\eps_{cc}=0.38$, $\eps_{bc}=0.18$, $\eps_{cb}=0.03$,
$\eps_{qb}=\eps_{gb}=0.03$, $\eps_{qc}=\eps_{gc}=0.32$, 
$\eps_{\tau\tau}=0.8$, $\eps_{\tau b}=\eps_{\tau c}=\eps_{\tau q}=0$,
where, for example, $\eps_{bb}$ ($\eps_{bc}$) is the probability
that a $b$-quark jet is tagged as a $b$ ($c$). Only events in which
the jets or $\tau$'s have $|\cos\theta|<0.94$ (corresponding
to a nose cone of $20^\circ$) are considered. A jet final state is deemed
to be: $b\anti b$ if one or more jets is tagged as a $b$; $c\anti c$
if no jet is tagged as a $b$, but one or more jets is tagged as a $c$;
and $q\anti q,gg$ if neither jet is tagged as a $b$ or a $c$.
Background and signal events are analyzed in exactly the same manner.
Note, in particular, that even though the $\pzero$ does not decay
to $c\anti c$, some of its $b\anti b$ and $gg$ decays will be identified
as $c\anti c$. 
In Fig.~\ref{figlum}, we plot the integrated luminosity required to
achieve $S_i/\sqrt {B_i}=5$ in a given channel, $i$ 
(as defined after tagging),
taking $\rts=\mpzero$. We also show the luminosity $L$ needed 
to achieve $\sum_k S_k/\sqrt{\sum_k B_k}=5$, where the optimal choice
of channels $k$ is determined for each $\mpzero$. We observe that
very modest $L$ is needed unless $\mpzero\sim\mz$.
\begin{table}
\caption{Luminosity (in units of $0.01\fbi$)
required to scan 
from $M_{\rm min}+(\mz-90)$ to $M_{\rm min}+(\mz-90)+5$ (GeV units)
and either discover or eliminate the $\pzero$ at the $3\sigma$
level. For scan details, see text.}
\begin{center}
\begin{tabular}{|c|c|c|c|c|c|c|c|c|c|c|}
\hline
 $M_{\rm min}$ & 11 & 16 & 21 & 26 &
 31 & 36 & 41 & 46 & 51 & 56\\
 $L$ & 0.028 & 0.051 & 0.079 & 0.10 & 0.13 & 0.18 & 
0.23 & 0.29 & 0.40 & 0.55\\
\hline
 $M_{\rm min}$  & 61 & 66 & 71 & 76 & 81
 & 86 & 91 & 96 & 101 & 106\\ 
 $L$ & 0.77 & 1.2 & 2.2 & 5.3 & 17 & 166
 & 274 & 52& 23 & 15\\
\hline
 $M_{\rm min}$ & 111 & 116 & 121 & 126
& 131 & 136 & 141 & 146 & 151 & 156 \\
 $L$  & 11 & 9.4 & 8.5 & 8.1 & 8.2 &
 8.2 & 8.3 & 8.7 & 8.9 & 9.0 \\
\hline
\end{tabular}
\end{center}
\label{lumtable}
\end{table}
Of course, if we do not have any information regarding the $\pzero$ mass,
we must scan for the resonance. To estimate
the luminosity required for scanning a given interval
so as to either discover or eliminate the $\pzero$, we have adopted
the following approach.  We imagine choosing $\rts$ values separated
by $2\sigrts$. We assume the worst case
scenario in which the resonance sits midway between the two $\rts$
values.  The signal and (separately) background rates for these two $\rts$
values are summed together (for the optimal channel combination)
and the net $N_{SD}\equiv (S_1+S_2)/(B_1+B_2)^{1/2}$ is computed. 
We require $N_{SD}=3$ to claim a signal. The luminosity required
for a successful scan of a given interval is computed 
assuming that the resonance lies between the {\it last} two scan points.
This, in combination with the fact that $\sigrts$ for $R=0.003\%$ is
typically a factor of two smaller than $\Gamma_{\rm tot}^{\pzero}$
(implying that points further away than $\sigrts$ from the resonance
could be usefully included in establishing a signal) will imply
that the integrated luminosities given below are quite conservative.
We give in Table~\ref{lumtable} the integrated luminosity for a $3\sigma$
$\pzero$ discovery after scanning the indicated $5\gev$ intervals, assuming
$\mpzero$ lies within that interval. If the $\pzero$ is as light
as expected in the extended BESS model, then
the prospects for discovery by scanning would be excellent. For example,
a $\pzero$ lying in the $\sim 10\gev$ to $\sim 76\gev$ mass interval
can be either discovered or eliminated at the $3\sigma$ level with
just $0.11\fbi$ of total luminosity, distributed in proportion to
the (combined) luminosity plotted in Fig.~\ref{figlum}. A $\pzero$
with $\mpzero\sim\mz$ would be much more difficult to discover
unless its mass was approximately known. A $3\sigma$ scan
of the mass interval from $\sim 106\gev$ to $161\gev$ would require
about $1\fbi$ of integrated luminosity.

\section*{Conclusion}

We have demonstrated the very substantial suitability, and in many respects
superiority, of a muon collider for exploring the full range of physics
associated with a strongly-interacting electroweak sector as typified
by the extended BESS model.  Especially interesting is the potential for
discovering any light pseudo-Nambu-Goldstone boson
with lepton couplings by scanning. Such bosons are
a general feature of models of a strongly interacting electroweak
sector and may prove to be quite difficult to detect in any other way.


\end{document}